\documentclass[11pt, a4paper]{article}

\usepackage{jheppub, slashed}

\newcommand{\ChS}{\text{CS}}

\newcommand{\gf}{\mathfrak{g}}

\newcommand{\del}{\partial}

\newcommand{\Ds}{\slashed{D}}

\newcommand{\vev}[1]{\langle #1 \rangle}

\renewcommand{\Im}{\mathop{\mathrm{Im}}\nolimits}
\renewcommand{\Re}{\mathop{\mathrm{Re}}\nolimits}

\newcommand{\Tr}{\mathop{\mathrm{Tr}}\nolimits}

\newcommand{\SU}{\mathrm{SU}}

\newcommand{\Spin}{\mathrm{Spin}}

\newcommand{\U}{\mathrm{U}}

\newcommand{\iso}{\cong}

\newcommand{\Z}{\mathbb{Z}}

\newcommand{\R}{\mathbb{R}}
\newcommand{\C}{\mathbb{C}}

\let\nc\newcommand
\let\renc\renewcommand
\nc{\wbar}{\overline}
\let\td\tilde
\let\wtd\widetilde
\let\wht\widehat
\let\mcl\mathcal

\nc{\ab}{{\bar{a}}} \nc{\at}{\tilde{a}} \nc{\ah}{\hat{a}}
\nc{\bb}{{\bar{b}}} \nc{\bt}{\tilde{b}} \nc{\bh}{\hat{b}}
\nc{\cb}{{\bar{c}}} \nc{\ct}{\tilde{c}} 
\nc{\db}{{\bar{d}}} \nc{\dt}{\tilde{d}} \renc{\dh}{\hat{d}}
\nc{\eb}{{\bar{e}}} \nc{\et}{\tilde{e}} \nc{\eh}{\hat{e}}
\nc{\fb}{{\bar{f}}} \nc{\ft}{\tilde{f}} \nc{\fh}{\hat{f}}
\nc{\gb}{{\bar{g}}} \nc{\gt}{\tilde{g}} \nc{\gh}{\hat{g}}
\nc{\hb}{{\bar{h}}} \nc{\hh}{\hat{h}} 
\nc{\ib}{{\bar{\imath}}} \nc{\ih}{\hat{\imath}} 
\nc{\jb}{{\bar{\jmath}}} \nc{\jt}{\tilde{\jmath}} \nc{\jh}{\hat{\jmath}}
\nc{\kb}{{\bar{k}}} \nc{\kt}{\tilde{k}} \nc{\kh}{\hat{k}}
\nc{\lb}{{\bar{l}}} \nc{\lt}{\tilde{l}} \nc{\lh}{\hat{l}}
\nc{\mb}{{\bar{m}}} \nc{\mt}{\tilde{m}} \nc{\mh}{\hat{m}}
\nc{\nb}{{\bar{n}}} \nc{\nt}{\tilde{n}} \nc{\nh}{\hat{n}}
\nc{\ob}{{\bar{o}}} \nc{\ot}{\tilde{o}} \nc{\oh}{\hat{o}}
\nc{\pb}{{\bar{p}}} \nc{\pt}{\tilde{p}} \nc{\ph}{\hat{p}}
\nc{\qb}{{\bar{q}}} \nc{\qt}{\tilde{q}} \nc{\qh}{\hat{q}}
\nc{\rb}{{\bar{r}}} \nc{\rt}{\tilde{r}} \nc{\rh}{\hat{r}}
\renc{\sb}{{\bar{s}}} \nc{\st}{\tilde{s}} \nc{\sh}{\hat{s}}
\nc{\tb}{{\bar{t}}} \renc{\th}{\hat{t}} 
\nc{\ub}{{\bar{u}}} \nc{\ut}{\tilde{u}} \nc{\uh}{\hat{u}}
\nc{\vb}{{\bar{v}}} \nc{\vt}{\tilde{v}} \nc{\vh}{\hat{v}}
\nc{\wb}{{\bar{w}}} \nc{\wt}{\tilde{w}} \nc{\wh}{\hat{w}}
\nc{\xb}{{\bar{x}}} \nc{\xt}{\tilde{x}} \nc{\xh}{\hat{x}}
\nc{\yb}{{\bar{y}}} \nc{\yt}{\tilde{y}} \nc{\yh}{\hat{y}}
\nc{\zb}{{\bar{z}}} \nc{\zt}{\tilde{z}} \nc{\zh}{\hat{z}}

\nc{\Ab}{\wbar{A}} \nc{\At}{\wtd{A}} \nc{\Ah}{\wht{A}}
\nc{\Bb}{\wbar{B}} \nc{\Bt}{\wtd{B}} \nc{\Bh}{\wht{B}}
\nc{\Cb}{\wbar{C}} \nc{\Ct}{\wtd{C}} \nc{\Ch}{\wht{C}}
\nc{\Db}{\wbar{D}} \nc{\Dt}{\wtd{D}} \nc{\Dh}{\wht{D}}
\nc{\Eb}{\wbar{E}} \nc{\Et}{\wtd{E}} \nc{\Eh}{\wht{E}}
\nc{\Fb}{\wbar{F}} \nc{\Ft}{\wtd{F}} \nc{\Fh}{\wht{F}}
\nc{\Gb}{\wbar{G}} \nc{\Gt}{\wtd{G}} \nc{\Gh}{\wht{G}}
\nc{\Hb}{\wbar{H}} \nc{\Ht}{\wtd{H}} \nc{\Hh}{\wht{H}}
\nc{\Ib}{\wbar{I}} \nc{\It}{\wtd{I}} \nc{\Ih}{\wht{I}}
\nc{\Jb}{\wbar{J}} \nc{\Jt}{\wtd{J}} \nc{\Jh}{\wht{J}}
\nc{\Kb}{\wbar{K}} \nc{\Kt}{\wtd{K}} \nc{\Kh}{\wht{K}}
\nc{\Lb}{\wbar{L}} \nc{\Lt}{\wtd{L}} \nc{\Lh}{\wht{L}}
\nc{\Mb}{\wbar{M}} \nc{\Mt}{\wtd{M}} \nc{\Mh}{\wht{M}}
\nc{\Nb}{\wbar{N}} \nc{\Nt}{\wtd{N}} \nc{\Nh}{\wht{N}}
\nc{\Ob}{\wbar{O}} \nc{\Ot}{\wtd{O}} \nc{\Oh}{\wht{O}}
\nc{\Pb}{\wbar{P}} \nc{\Pt}{\wtd{P}} \nc{\Ph}{\wht{P}}
\nc{\Qb}{\wbar{Q}} \nc{\Qt}{\wtd{Q}} \nc{\Qh}{\wht{Q}}
\nc{\Rb}{\wbar{R}} \nc{\Rt}{\wtd{R}} \nc{\Rh}{\wht{R}}
\nc{\Sb}{\wbar{S}} \nc{\St}{\wtd{S}} \nc{\Sh}{\wht{S}}
\nc{\Tb}{\wbar{T}} \nc{\Tt}{\wtd{T}} \nc{\Th}{\wht{T}}
\nc{\Ub}{\wbar{U}} \nc{\Ut}{\wtd{U}} \nc{\Uh}{\wht{U}}
\nc{\Vb}{\wbar{V}} \nc{\Vt}{\wtd{V}} \nc{\Vh}{\wht{V}}
\nc{\Wb}{\wbar{W}} \nc{\Wt}{\wtd{W}} \nc{\Wh}{\wht{W}}
\nc{\Xb}{\wbar{X}} \nc{\Xt}{\wtd{X}} \nc{\Xh}{\wht{X}}
\nc{\Yb}{\wbar{Y}} \nc{\Yt}{\wtd{Y}} \nc{\Yh}{\wht{Y}}
\nc{\Zb}{\wbar{Z}} \nc{\Zt}{\wtd{Z}} \nc{\Zh}{\wht{Z}}

\nc{\CA}{\mcl{A}} \nc{\CAb}{\wbar{\CA}} \nc{\CAt}{\wtd{\CA}} \nc{\CAh}{\wht{\CA}}
\nc{\CB}{\mcl{B}} \nc{\CBb}{\wbar{\CB}} \nc{\CBt}{\wtd{\CB}} \nc{\CBh}{\wht{\CB}}
\nc{\CC}{\mcl{C}} \nc{\CCb}{\wbar{\CC}} \nc{\CCt}{\wtd{\CC}} \nc{\CCh}{\wht{\CC}}
\nc{\cD}{\mcl{D}} \nc{\cDb}{\wbar{\cD}} \nc{\cDt}{\wtd{\cC}} \nc{\cDh}{\wht{\cD}}
\nc{\CE}{\mcl{E}} \nc{\CEb}{\wbar{\CE}} \nc{\CEt}{\wtd{\CE}} \nc{\CEh}{\wht{\CE}}
\nc{\CF}{\mcl{F}} \nc{\CFb}{\wbar{\CF}} \nc{\CFt}{\wtd{\CF}} \nc{\CFh}{\wht{\CF}}
\nc{\CG}{\mcl{G}} \nc{\CGb}{\wbar{\CG}} \nc{\CGt}{\wtd{\CG}} \nc{\CGh}{\wht{\CG}}
\nc{\CH}{\mcl{H}} \nc{\CHb}{\wbar{\CH}} \nc{\CHt}{\wtd{\CH}} \nc{\CHh}{\wht{\CH}}
\nc{\CI}{\mcl{I}} \nc{\CIb}{\wbar{\CI}} \nc{\CIt}{\wtd{\CI}} \nc{\CIh}{\wht{\CI}}
\nc{\CJ}{\mcl{J}} \nc{\CJb}{\wbar{\CJ}} \nc{\CJt}{\wtd{\CJ}} \nc{\CJh}{\wht{\CJ}}
\nc{\CK}{\mcl{K}} \nc{\CKb}{\wbar{\CK}} \nc{\CKt}{\wtd{\CK}} \nc{\CKh}{\wht{\CK}}
\nc{\CL}{\mcl{L}} \nc{\CLb}{\wbar{\CL}} \nc{\CLt}{\wtd{\CL}} \nc{\CLh}{\wht{\CL}}
\nc{\CM}{\mcl{M}} \nc{\CMb}{\wbar{\CM}} \nc{\CMt}{\wtd{\CM}} \nc{\CMh}{\wht{\CM}}
\nc{\CN}{\mcl{N}} \nc{\CNb}{\wbar{\CN}} \nc{\CNt}{\wtd{\CN}} \nc{\CNh}{\wht{\CN}}
\nc{\CO}{\mcl{O}} \nc{\COb}{\wbar{\CO}} \nc{\COt}{\wtd{\CO}} \nc{\COh}{\wht{\CO}}
\nc{\CQ}{\mcl{Q}} \nc{\CQb}{\wbar{\CQ}} \nc{\CQt}{\wtd{\CQ}} \nc{\CQh}{\wht{\CQ}}
\nc{\CR}{\mcl{R}} \nc{\CRb}{\wbar{\CR}} \nc{\CRt}{\wtd{\CR}} \nc{\CRh}{\wht{\CR}}
\nc{\CS}{\mcl{S}} \nc{\CSb}{\wbar{\CS}} \nc{\CSt}{\wtd{\CS}} \nc{\CSh}{\wht{\CS}}
\nc{\CT}{\mcl{T}} \nc{\CTb}{\wbar{\CT}} \nc{\CTt}{\wtd{\CT}} \nc{\CTh}{\wht{\CT}}
\nc{\CU}{\mcl{U}} \nc{\CUb}{\wbar{\CU}} \nc{\CUt}{\wtd{\CU}} \nc{\CUh}{\wht{\CU}}
\nc{\CV}{\mcl{V}} \nc{\CVb}{\wbar{\CV}} \nc{\CVt}{\wtd{\CV}} \nc{\CVh}{\wht{\CV}}
\nc{\CW}{\mcl{W}} \nc{\CWb}{\wbar{\CW}} \nc{\CWt}{\wtd{\CW}} \nc{\CWh}{\wht{\CW}}
\nc{\CX}{\mcl{X}} \nc{\CXb}{\wbar{\CX}} \nc{\CXt}{\wtd{\CX}} \nc{\CXh}{\wht{\CX}}
\nc{\CY}{\mcl{Y}} \nc{\CYb}{\wbar{\CY}} \nc{\CYt}{\wtd{\CY}} \nc{\CYh}{\wht{\CY}}
\nc{\CZ}{\mcl{Z}} \nc{\CZb}{\wbar{\CZ}} \nc{\CZt}{\wtd{\CZ}} \nc{\CZh}{\wht{\CZ}}

\let\eps\epsilon
\let\ups\upsilon
\let\veps\varepsilon
\let\vtht\vartheta
\let\vsgm\varsigma
\let\vphi\varphi
\let\vrho\varrho

\nc{\alphab}{\bar{\alpha}} \nc{\alphat}{\td{\alpha}} \nc{\alphah}{\hat{\alpha}}
\nc{\betab}{\bar{\beta}}   \nc{\betat}{\td{\beta}}   \nc{\betah}{\hat{\beta}} 
\nc{\gammab}{\bar{\gamma}} \nc{\gammat}{\td{\gamma}} \nc{\gammah}{\hat{\gamma}} 
\nc{\deltab}{\bar{\delta}} \nc{\deltat}{\td{\delta}} \nc{\deltah}{\hat{\delta}} 
\nc{\epsilonb}{\bar{\eps}} \nc{\epsilont}{\td{\eps}} \nc{\epsilonh}{\hat{\eps}} 
\nc{\vepsb}{\bar{\veps}}   \nc{\vepst}{\td{\veps}}   \nc{\vepsh}{\hat{\veps}} 
\nc{\zetab}{\bar{\zeta}}   \nc{\zetat}{\td{\zeta}}   \nc{\zetah}{\hat{\zeta}} 
\nc{\etab}{\bar{\eta}}     \nc{\etat}{\td{\eta}}     \nc{\etah}{\hat{\eta}} 
\nc{\thetab}{\bar{\theta}} \nc{\thetat}{\td{\theta}} \nc{\thetah}{\hat{\theta}} 
\nc{\vthetab}{\bar{\vtht}} \nc{\vthetat}{\td{\vtht}} \nc{\vthetah}{\hat{\vtht}} 
\nc{\lambdab}{\bar{\lambda}} \nc{\lambdat}{\td{\lambda}} \nc{\lambdah}{\hat{\lambda}} 
\nc{\iotab}{\bar{\iota}}   \nc{\iotat}{\td{\iota}}   \nc{\iotah}{\hat{\iota}} 
\nc{\kappab}{\bar{\kappa}} \nc{\kappat}{\td{\kappa}} \nc{\kappah}{\hat{\kappa}} 
\nc{\lmdb}{\bar{\lmd}}     \nc{\lmdt}{\td{\lmd}}     \nc{\lmdh}{\hat{\lmd}} 
\nc{\mub}{\bar{\mu}}       \nc{\mut}{\td{\mu}}       \nc{\muh}{\hat{\mu}} 
\nc{\nub}{\bar{\nu}}       \nc{\nut}{\td{\nu}}       \nc{\nuh}{\hat{\nu}} 
\nc{\xib}{\bar{\xi}}       \nc{\xit}{\td{\xi}}       \nc{\xih}{\hat{\xi}} 
\nc{\pib}{\bar{\pi}}       \nc{\pit}{\td{\pi}}       \nc{\pih}{\hat{\pi}} 
\nc{\vpib}{\bar{\vpi}}     \nc{\vpit}{\td{\vpi}}     \nc{\vpih}{\hat{\vpi}} 
\nc{\rhob}{\bar{\rho}}     \nc{\rhot}{\td{\rho}}     \nc{\rhoh}{\hat{\rho}} 
\nc{\vrhob}{\bar{\vrho}}   \nc{\vrhot}{\td{\vrho}}   \nc{\vrhoh}{\hat{\vrho}} 
\nc{\sigmab}{\bar{\sigma}} \nc{\sigmat}{\td{\sigma}} \nc{\sigmah}{\hat{\sigma}} 
\nc{\vsigmab}{\bar{\vsgm}} \nc{\vsigmat}{\td{\vsgm}} \nc{\vsigmah}{\hat{\vsgm}} 
\nc{\taub}{\bar{\tau}}     \nc{\taut}{\td{\tau}}     \nc{\tauh}{\hat{\tau}} 
\nc{\upsilonb}{\bar{\ups}} \nc{\upsilont}{\td{\ups}} \nc{\upsilonh}{\hat{\ups}} 
\nc{\phib}{\bar{\phi}}     \nc{\phit}{\td{\phi}}     \nc{\phih}{\hat{\phi}} 
\nc{\varphib}{\bar{\vphi}}   \nc{\varphit}{\td{\vphi}}   \nc{\varphih}{\hat{\vphi}} 
\nc{\chib}{\bar{\chi}}     \nc{\chit}{\td{\chi}}     \nc{\chih}{\hat{\chi}} 
\nc{\psib}{\bar{\psi}}     \nc{\psit}{\td{\psi}}     \nc{\psih}{\hat{\psi}} 
\nc{\omegab}{\bar{\omega}} \nc{\omegat}{\td{\omega}} \nc{\omegah}{\hat{\omega}} 

\nc{\Gammab}{\wbar{\Gamma}}     \nc{\Gammat}{\wtd{\Gamma}}     \nc{\Gammah}{\wht{\Gamma}}
\nc{\Deltab}{\wbar{\Delta}}     \nc{\Deltat}{\wtd{\Delta}}     \nc{\Deltah}{\wht{\Delta}}
\nc{\Thetab}{\wbar{\Theta}}     \nc{\Thetat}{\wtd{\Theta}}     \nc{\Thetah}{\wht{\Theta}}
\nc{\Lambdab}{\wbar{\Lambda}}   \nc{\Lambdat}{\wtd{\Lambda}}   \nc{\Lambdah}{\wht{\Lambda}}
\nc{\Xib}{\wbar{\Xi}}           \nc{\Xit}{\wtd{\Xi}}           \nc{\Xih}{\wht{\Xi}}
\nc{\Pib}{\wbar{\Pi}}           \nc{\Pit}{\wtd{\Pi}}           \nc{\Pih}{\wht{\Pi}}
\nc{\Sigmab}{\wbar{\Sigma}}     \nc{\Sigmat}{\wtd{\Sigma}}     \nc{\Sigmah}{\wht{\Sigma}}
\nc{\Upsilonb}{\wbar{\Upsilon}} \nc{\Upsilont}{\wtd{\Upsilon}} \nc{\Upsilonh}{\wht{\Upsilon}}
\nc{\Phib}{\wbar{\Phi}}         \nc{\Phit}{\wtd{\Phi}}         \nc{\Phih}{\wht{\Phi}}
\nc{\Psib}{\wbar{\Psi}}         \nc{\Psit}{\wtd{\Psi}}         \nc{\Psih}{\wht{\Psi}}
\nc{\Omegab}{\wbar{\Omega}}     \nc{\Omegat}{\wtd{\Omega}}     \nc{\Omegah}{\wht{\Omega}}

\newcommand{\upsb}{\bar{\ups}}
\newcommand{\Tbb}{\mathbb{T}}
\newcommand{\rmd}{\mathrm{d}}
\newcommand{\nablas}{\slashed{\nabla}}

\title{3d TQFT from 6d SCFT}

\author{Junya Yagi}

\affiliation{Department of Physics, National University of Singapore, \\
  2 Science Drive 3, Singapore 117551}

\emailAdd{yagi@nus.edu.sg}

\abstract{We study the six-dimensional $(2,0)$ superconformal field
  theory on $S^1 \times S^2 \times M$ via compactification to five
  dimensions, where $M$ is a three-manifold.  Twisted along $M$, the
  five-dimensional theory has a half of $\CN = (2,2)$ supersymmetry on
  $S^2$, the other half being broken by a superpotential.  We show
  that in the limit where $M$ is infinitely large, the twisted theory
  reduces to a three-dimensional topological quantum field theory
  which is closely related to Chern-Simons theory for the
  complexified gauge group.}

\keywords{Extended Supersymmetry, Field Theories in Higher Dimensions,
  Topological Field Theories}

\arxivnumber{1305.0291}

\renewcommand{\ell}{l}

\begin{document}
\maketitle
\flushbottom

\section{Introduction}

In the past several years various correspondences have been discovered
between supersymmetric theories in $d$ dimensions and
nonsupersymmetric theories in $6 - d$ dimensions.  The most famous
example is the AGT correspondence \cite{Alday:2009aq} between $\CN =
2$ supersymmetric gauge theories in four dimensions and Liouville (or
more generally, Toda \cite{Mironov:2009by, Wyllard:2009hg,
  Taki:2009zd}) conformal field theory, which has now been extensively
investigated.  A less studied example is the 3d/3d correspondence
\cite{Dimofte:2011py} between $\CN = 2$ superconformal field theories
(SCFTs) in three dimensions and Chern-Simons theory with complex
gauge group.  The aim of the present paper is to gain a deeper
understanding of the 3d/3d correspondence, via five-dimensional
maximally supersymmetric Yang-Mills theory (5d MSYM).

It has been argued that these intriguing correspondences originate
from the $\CN = (2,0)$ SCFT in six dimensions, formulated on the
product $X \times M$ of a $d$-dimensional space $X$ and a
$(6-d)$-dimensional space $M$.  Typically, one allows $M$ to be a
fairly general manifold and twists the theory along $M$, but chooses
$X$ to be very specific so that the theory admits supersymmetry
without twisting along $X$.  On the one hand, compactification of the
theory on $M$ produces a supersymmetric theory $T[M]$ on $X$.  On the
other hand, upon ``supersymmetric localization on $X$,'' the theory
reduces to a nonsupersymmetric theory $\Tbb[X]$ on $M$.  Identifying
protected quantities after these different procedures, one establishes
a correspondence between $T[M]$ and $\Tbb[X]$.

The AGT correspondence arises by taking $X = S^4$ (either round or
squashed \cite{Hama:2012bg}) and $M$ to be punctured Riemann surfaces;
$T[M]$ is an $\CN = 2$ theory of class $\CS$ \cite{Gaiotto:2009we,
  Gaiotto:2009hg} and $\Tbb[X]$ is Toda theory.  In one version of
3d/3d correspondence, one takes $X = S^3$ and $M$ to be
three-manifolds.  Then $T[M]$ is an $\CN = 2$ SCFT and $\Tbb[X]$ is
analytically continued Chern-Simons theory \cite{Terashima:2011qi,
  Dimofte:2011jd, Dimofte:2011ju}.

As one can see from these examples, the theory $\Tbb[X]$ is rather
rigid in the sense that it has few free parameters, reflecting the
highly constrained geometry of $X$.  Furthermore, it is often
conformal or even topological since the twist along $M$ decouples some
part or the whole of the dependency on the metric chosen on $M$ to
formulate the $(2,0)$ theory, by making the corresponding components
of the stress tensor $Q$-exact with respect to some supercharge $Q$.
Because of this decoupling, quantities involving only $Q$-invariant
operators and states are protected under the compactification
described above.  These are also the quantities captured by the
localization method.

For the 3d/3d correspondence of our interest, one takes
\begin{equation}
  X = S^1 \times S^2,
\end{equation}
with $S^1$ a circle and $S^2$ a round two-sphere.  On $S^1 \times
S^2$, the $\CN = 2$ SCFT $T[M]$ has a class of correlation functions
that take the form of an index.  This class in particular includes the
superconformal index, which is the partition function with a twisted
boundary condition as the fields go around the $S^1$.%
\footnote{Boundary conditions relevant in the present context are
  given by the combined action of an R-symmetry rotation and flavor
  symmetry transformations, which preserve a half of the eight
  supersymmetries of the $\CN = 2$ superconformal group on $\R \times
  S^2$ \cite{Imamura:2011su}.  By dimensional reduction the four
  preserved supersymmetries generate $\CN = (2,2)$ supersymmetry on
  $S^2$.}
Being indices, these correlators are invariant under continuous
changes in the parameters of the theory.  This suggests that they come
from six-dimensional indices that are protected under deformations of
the geometry of $M$.  So these should be the quantities $\Tbb[X]$ on
$M$ captures, and we deduce that $\Tbb[X]$ is a topological quantum
field theory (TQFT).  The expectation is that $\Tbb[X]$ is complex
Chern-Simons theory \cite{Dimofte:2011py}.

While the geometric structures that emerge when $M$ is varied provide
important hints for identifying $T[M]$, the rigidity of $X$ prevents
one from learning about $\Tbb[X]$ in a similar manner.  So far the
identification of $\Tbb[X]$ has mainly been achieved through more
indirect routes, for example, by looking for structures in the
partition function of $T[M]$ on $X$, or by relating the setup to
another case where the correspondence has already been established.
Clearly it is desirable to find a more direct derivation of $\Tbb[X]$
starting from six dimensions and following the above line of
reasoning.  Our goal is to actually do this for $X = S^1 \times S^2$.

There is an obvious obstacle, however: localization computations
require a Lagrangian description, but the $(2,0)$ theory has none
known.

Happily, the problem can be avoided in our case.  Since $X$ contains
an $S^1$ factor, one can first consider compactifying the $(2,0)$
theory on the $S^1$ down to five dimensions.  This gives 5d MSYM on
$S^2 \times M$ with gauge coupling
\begin{equation}
  \label{e2R}
  e^2 \propto R,
\end{equation}
where $R$ is the radius of the $S^1$.  The latter theory does have a
Lagrangian, so one can take this as a starting point and hope to
derive $\Tbb[X]$ by localization of the path integral.  A similar
strategy was employed in \cite{Kawano:2012up, Fukuda:2012jr} to
approach the conjecture \cite{Gadde:2009kb, Gadde:2011ik} that
$\Tbb[S^1 \times S^3]$ is the zero area limit of $q$-deformed
Yang-Mills theory in two dimensions.

With this in mind, in section~\ref{MSYM} we construct 5d MSYM on $S^2
\times M$, twisted along $M$ in the way that descends from the
aforementioned twist in six dimensions.  The idea is to look at the
dimensional reduction of the theory on $M$.  An analysis of the
twisting reveals that this is an $\CN = (2,2)$ supersymmetric gauge
theory on $S^2$ with adjoint chiral multiplets, for which we know the
general form of Lagrangian \cite{Benini:2012ui, Doroud:2012xw}.  Then
we lift this Lagrangian to five dimensions.  This step can be done
straightforwardly, except that we have to determine the superpotential
which reproduces the standard Lagrangian on $\R^5$ after we set $M =
\R^3$ and replace the $S^2$ metric by the flat $\R^2$ metric.  The
required superpotential is given by the Chern-Simons functional for a
complexified gauge field $\CA$ on $M$.

This superpotential violates the vector R-symmetry and, as a
consequence, breaks $\CN = (2,2)$ supersymmetry on $S^2$.  It is
therefore not possible to realize on $S^2 \times M$ all of the four
supersymmetries.  We will see, however, that a half of them can still
be preserved by adding appropriate correction terms.  That the
curvature of the $S^2$ halves the number of supersymmetries is not
totally surprising; if one instead tried to realize the theory on a
stack of D4-branes wrapped on $S^2 \times M \subset T^*S^2 \times
T^*M$, then one would get a slightly different situation where the
theory is twisted along both the $S^2$ and $M$, and the twisting along
the $S^2$ would also halve the number.

To connect to the story of the 3d/3d correspondence, we propose that
the $(2,0)$ theory can be formulated on $S^1 \times S^2 \times M$ in
such a way that when the $S^1$ is small, it reduces at low energies to
the twisted 5d MSYM on $S^2 \times M$ constructed as above, including
these corrections to the F-term.  One piece of evidence is that the
five-dimensional theory is invariant under the simultaneous rescaling
of the coupling $e^2$, the radius $r$ of the $S^2$, and the length
scale of $M$.  In view of the relation \eqref{e2R}, this is nothing
but the scale invariance of the $(2,0)$ theory.

From the consideration on the nature of protected quantities in six
dimensions, we expect that the twisted theory on $S^2 \times M$
becomes topological along $M$ in the limit $R \to 0$ where the $(2,0)$
theory reduces to 5d MSYM.  Actually, because of the relation
\eqref{e2R}, this limit is the zero coupling limit and the physics is
not very interesting there.  So we instead consider the limit where
$R$ and $r$ are both sent to zero with $r/R$ kept fixed, or
equivalently, the metric $g_M$ on $M$ is rescaled by an infinitely
large factor.  Since the twisted $(2,0)$ theory is presumably
topological on $M$, we can do this without affecting the protected
quantities.

In section \ref{TFT}, we show that the $g_M$-dependence of the twisted
theory indeed decouples in the $Q$-invariant sector when $M$ becomes
infinitely large, with $Q$ being one of the two preserved
supercharges.  Therefore, the infinite volume limit of the twisted
theory is a TQFT on $M$, and we identify it with $\Tbb[X]$.%
\footnote{In fact, the choice of $Q$ used in ``localizing'' the
  $(2,0)$ theory may not correspond to $Q$ we use for the twisted
  theory.  In particular, one can choose it to be one of the
  supercharges restored in the infinite volume limit.  The difference
  does not matter as far as the superconformal index is concerned,
  since the twisted boundary condition can be chosen to be invariant
  under either supercharge.}
Also in this limit the vector R-symmetry and the full $\CN = (2,2)$
supersymmetry are restored.  This is an expected behavior since the
$(2,0)$ theory on $S^1 \times S^2 \times M$ must preserve four
supersymmetries in order for $T[M]$ to have $\CN = 2$ supersymmetry.

Encouraged by these findings, in section~\ref{CS} we attempt to
determine $\Tbb[X]$.  By suitably deforming the action in the infinite
volume limit, we localize the path integral and express it as the sum
of path integrals labeled by the magnetic flux $B$ through the $S^2$.
If $G$ is the gauge group of the twisted theory and $G_B \subset G$ is
the stabilizer of $B$ (the subgroup consisting of all $g \in G$ such
that $g^{-1}Bg = B$), then each summand is the path integral
\linebreak for a
variant of Chern-Simons theory on $M$ whose gauge group is the
complexification $(G_B)_\C$ of $G_B$:
\begin{equation}
  \sum_B \int_{\CM^{\text{ss}}_B} \widetilde{\cD\CA_B}
  \exp\bigl(iS_{\ChS}(\CA_B)\bigr).
\end{equation}
We find the Chern-Simons level $k = 0$, which is the correct value
for the 3d/3d correspondence.  The differences from the ordinary
complex Chern-Simons theory are that the measure is modified by
one-loop contributions, and that the integration is performed over the
space $\CM_B^{\text{ss}}$ of semistable $(G_B)_\C$-orbits in the space
of complex gauge fields, rather than all $(G_B)_\C$-orbits.

Unfortunately, in this work we will not be able to write down an
explicit formula for the modified measure.  The difficulty lies in
that we cannot compute the one-loop determinants explicitly, as we
keep $M$ to be a general three-manifold.  We hope to address this
issue in future research.

\section{5d MSYM on \texorpdfstring{\boldmath{$S^2 \times M$}}{S**2 x M}}
\label{MSYM}

From now on we forget about the 3d/3d correspondence and the
six-dimensional physics underlying it.  Our goal here is to formulate
5d MSYM on $S^2 \times M$, where $S^2$ is a two-sphere equipped with
the round metric $h$ of radius $r$ and $M$ is a three-manifold with
metric $g_M$.  The metric of the total product space is $g = h \oplus
g_M$.  Our conventions for spinors on $S^2$ are summarized in appendix
\ref{S2}.

The theory consists of a gauge field $A$, five adjoint scalars $X^I$
($I = 1$, $\dotsc$, $5$), and sixteen adjoint spinors $\Psi$.  The
R-symmetry group of the theory is $\Spin(5)_R$.  Suppose that the
theory is formulated on a generic spin five-manifold $Y$, for which
the structure group of the spin bundle is $\Spin(5)_Y$.  Under
$\Spin(5)_Y \times \Spin(5)_R$, the fields transform as follows:
\begin{equation}
  \begin{aligned}
    A&\colon (\mathbf{5}, \mathbf{1}), \\
    X&\colon (\mathbf{1}, \mathbf{5}), \\
    \Psi&\colon (\mathbf{4}, \mathbf{4}).
  \end{aligned}
\end{equation}
Here $\mathbf{4}$ is a spinor representation of $\Spin(5)$.  On the
flat space $\R^5$, the theory has sixteen supersymmetries whose
generators transform as $(\mathbf{4}, \mathbf{4})$.

We are interested in the case when $Y = S^2 \times M$.  In this case
the supersymmetries \linebreak are completely broken generically, but we can
preserve a fraction of them with a suitable twist along $M$.

The relevant twist is performed in the following way.  On $S^2 \times
M$, the structure group of the spin bundle can be reduced to
\begin{equation}
  \Spin(2)_{S^2} \times \Spin(3)_M \iso \U(1)_{S^2} \times \SU(2)_M.
\end{equation}
We want to replace $\SU(2)_M$ by a different group $\SU(2)_M'$ that
acts trivially on some of the supercharges.  To this end we split
$\Spin(5)_R$ as
\begin{equation}
  \Spin(2)_R \times \Spin(3)_R \iso \U(1)_R \times \SU(2)_R,
\end{equation}
and take $\SU(2)_M'$ to be the diagonal subgroup of $\SU(2)_M \times
\SU(2)_R$.
Under $\SU(2)_M' \times \U(1)_{S^2} \times \U(1)_R$, the fields transform
as
\begin{equation}
  \begin{aligned}
    A&\colon \mathbf{1}^{(\pm 2, 0)}\oplus \mathbf{3}^{(0, 0)}, \\
    X&\colon \mathbf{1}^{(0,\pm 2)}\oplus \mathbf{3}^{(0, 0)}, \\
    \Psi&\colon \mathbf{1}^{(\pm 1, \pm 1)}\oplus \mathbf{3}^{(\pm 1, \pm 1)}.
  \end{aligned}
\end{equation}
We see that after the twisting four components of $\Psi$ become
scalars on $M$, hence so do four supercharges.

As a preliminary step to placing this twisted theory on $S^2 \times
M$, let us consider the situation where the $S^2$ is replaced by
$\R^2$.  On $\R^2 \times M$, the twisted theory has four
supersymmetries generated by the supercharges that are scalars on $M$.
Along the $\R^2$, two of these supercharges are spinors of chirality
positive and the other two negative.  Thus the twisted theory has $\CN
= (2,2)$ supersymmetry from the two-dimensional point of view.

We can readily figure out the field content of this $\CN = (2,2)$
supersymmetric theory.  There are two kinds of fields.

First, we have those that are scalars on $M$.  These are the
components $A_\mu$ of the gauge field along the $\R^2$, two real
scalars $\sigma_1$, $\sigma_2$ or one complex scalar $\sigma =
\sigma_1 - i\sigma_2$, and two Dirac fermions $\lambda$, $\lambdab$.
Supplemented with a real auxiliary field $D$, they form a vector
multiplet:
\begin{equation}
  (A_\mu, \sigma, \lambda, \lambdab, D).
\end{equation}

Second, we have those that are one-forms on $M$.  These are the
components $A_m$ of \linebreak the gauge field along $M$ and the remaining
scalars $X_m$ which combine into a complex \linebreak gauge field
\begin{equation}
  \CA_m = A_m + iX_m,
\end{equation}
and the remaining fermions $\psi_m$.  Together with complex
auxiliary fields $F_m$, they form chiral multiplets:
\begin{equation}
  (\CA_m, \psi_m, F_m).
\end{equation}

Therefore, upon dimensional reduction on $M$, the twisted theory
becomes an $\CN = (2,2)$ gauge theory with three adjoint chiral
multiplets.  The $\U(1)_R$ symmetry rotates the vector multiplet
scalars, and is identified with the axial R-symmetry $\U(1)_A$.

Now we want to place the theory on $S^2 \times M$, without twisting it
any further.  The dimensional reduction of this system on $M$ would
give an $\CN = (2,2)$ gauge theory on $S^2$.  Here we briefly review
the basic features of $\CN = (2,2)$ supersymmetry on $S^2$; more
details can be found in appendix \ref{SUSY}.

$\CN = (2,2)$ supersymmetry transformations on $S^2$ are parametrized
by a pair of conformal Killing spinors $(\veps, \vepsb)$ obeying the
equations
\begin{equation}
  \label{CKS}
    \nabla_\mu\veps = +\frac{1}{2r}\gamma_\mu\gamma_{\hat3}\veps, \quad
    \nabla_\mu\vepsb = -\frac{1}{2r}\gamma_\mu\gamma_{\hat3}\vepsb.
\end{equation}
We take $\veps$, $\vepsb$ to be commuting spinors.  Then supersymmetry
transformations are anticommuting.

Each of the above equations has two independent solutions,
$\veps_\alpha$ or $\vepsb_\alpha$ ($\alpha = 1$, $2$), so we have four
supercharges in total:
\begin{equation}
  \Qb_\alpha, \, Q_\alpha.
\end{equation}
They generate the supersymmetry transformations with $(\veps, \vepsb)
= (\veps_\alpha, 0)$ and $(0, \vepsb_\alpha)$, respectively.  We can
choose the parameters in such a way that
\begin{equation}
  \label{eebar}
  \vepsb_\alpha = \gamma_{\hat 3} \veps_\alpha
\end{equation}
and $\veps_1\veps_2 = -\vepsb_1\vepsb_2 = 1$.
The vector R-symmetry $\U(1)_V$ rotates $\Qb_\alpha$ and $Q_\alpha$ by
opposite phases.  Our normalization of the vector R-charge is that
$\Qb_\alpha$ has charge $q = +1$ and $Q_\alpha$ has $q = -1$.  As we
will see shortly, $\U(1)_V$ is actually broken to $\Z_2$ in our
theory.

The commutator of two supersymmetry transformations generates bosonic
continuous symmetries, which are $\U(1)_V$, rotations of $S^2$, and
gauge transformations.  Since none of these has $q = \pm 2$, it
follows immediately that
\begin{equation}
  \{\Qb_\alpha, \Qb_\beta\} = \{Q_\alpha, Q_\beta\} = 0.
\end{equation}
These relations will be useful later when we analyze the topological
property of the twisted theory.

To write down the supersymmetry transformation rules and the
Lagrangian for the twisted theory, we can lift their two-dimensional
counterparts to five dimensions.  This is done by promoting the fields
from functions on $S^2$ to those on $S^2 \times M$, and replacing
\begin{equation}
  A_m \to iD_m, \quad
  [A_m, A_n] \to iF_{mn},
\end{equation}
where $D_m = \nabla_m - iA_m$ are the covariant derivatives along $M$.
(For the purpose of writing down the five-dimensional Lagrangian, we
can assume that the gauge group is nonabelian.)  This lifting
operation commutes with any derivation commuting with $\nabla$,
especially supersymmetry transformations.  The lift of a
supersymmetric action is therefore invariant under the lifted
supersymmetry.

Following this procedure, we find that the supersymmetry
transformation of the vector multiplet is
\begin{equation}
  \label{VSUSY}
  \begin{aligned}
   \delta A_\mu &= -\frac{i}{2} (\vepsb\gamma_\mu\lambda
                                + \veps\gamma_\mu\lambdab), \\
   \delta\sigma &= \vepsb\gamma_-\lambda - \veps\gamma_+\lambdab, \\
   \delta\sigmab &= \vepsb\gamma_+\lambda - \veps\gamma_-\lambdab, \\
   \delta\lambda
   &= i\Bigl(F_{\hat1\hat2} \gamma_{\hat3}
            + \gamma_-\Ds\sigma
            + \gamma_+\Ds\sigmab
            + \frac{1}{2} [\sigma, \sigmab] \gamma_{\hat3} + iD\Bigr)\veps
            - \frac{i}{r} (\sigma\gamma_- - \sigmab\gamma_+)\veps, \\
   \delta\lambdab
   &= i\Bigl(F_{\hat1\hat2} \gamma_{\hat3}
             - \gamma_+\Ds\sigma
             - \gamma_-\Ds\sigmab
            - \frac{1}{2} [\sigma, \sigmab] \gamma_{\hat3} - iD\Bigr)\vepsb
            + \frac{i}{r} (\sigma\gamma_+ - \sigmab\gamma_-) \vepsb, \\
   \delta D &= -\frac{i}{2}\vepsb
               (\Ds\lambda
                + [\sigma, \gamma_+\lambda]
                + [\sigmab, \gamma_-\lambda])
               + \frac{i}{2r} \lambda\gamma_{\hat3}\vepsb
               \\ &\quad
               + \frac{i}{2}\veps
                  (\Ds\lambdab
                  - [\sigma, \gamma_-\lambda]
                  - [\sigmab, \gamma_+\lambda])
               - \frac{i}{2r} \lambdab\gamma_{\hat3}\veps.
\end{aligned}
\end{equation}

The complex gauge field $\CA_m$ has $q = 0$, for otherwise $\U(1)_V$
would not commute with the gauge symmetry which acts on $\CA_m$
inhomogeneously.  Then the supersymmetry transformation of the chiral
multiplets is
\begin{equation}
  \label{CSUSY}
  \begin{aligned}
    \delta\CA_m &= \vepsb\psi_m, \\
    \delta\CAb_m &= \veps\psib_m, \\
    \delta\psi_m &= i(\CF_{\mu m} \gamma^\mu
                        - i\cD_m\sigma\gamma_+ - i\cD_m\sigmab\gamma_-)\veps
                     + F_m \vepsb, \\
    \delta\psib_m &= i(\CFb_{\mu m} \gamma^\mu
                        + i\cDb_m\sigma\gamma_- + i\cDb_m\sigmab\gamma_+)\vepsb
                     + \Fb_m \veps, \\
   \delta F_m &=  i\veps(\Ds\psi_m
                          - [\sigma, \gamma_-\psi_m] - [\sigmab, \gamma_+\psi_m]
                          + i\cD_m\lambda), \\
                          \delta \Fb_m &= i\vepsb(\Ds\psib_m
                            + [\sigma, \gamma_+\psi_m] + [\sigmab, \gamma_-\psi_m]
                     + i\cDb_m\lambdab).
  \end{aligned}
\end{equation}
Here $\cD_m = \nabla_m -i\CA_m$ and $\CF_{\mu m} = \del_\mu\CA_m
- \del_m A_\mu -i[A_\mu, \CA_m]$; $\cDb_m$ and $\CFb_{\mu m}$ are
defined by similar formulas with $\CA_m$ replaced by $\CAb_m =
A_m - i X_m$.

Next, we lift the two-dimensional Lagrangian to five dimensions.  A
standard $\CN = (2,2)$ supersymmetric action on $S^2$ takes the form
\begin{equation}
  S = S_V + S_C + S_W,
\end{equation}
where $S_V$ and $S_C$ are the vector and chiral multiplet actions, and
$S_W$ is an F-term.  It is straightforward to write down the
five-dimensional versions of the first two; the vector multiplet
action is
\begin{multline}
  \label{SV-5d}
    S_V = \frac{1}{2e^2} \int \! \sqrt{g} \, \rmd^5x \Tr \Bigl[
            \Bigl(F_{\hat1\hat2} + \frac{\sigma_1}{r}\Bigr)^2
           + D^\mu\sigmab D_\mu\sigma
           + \frac{1}{4} [\sigma, \sigmab]^2 \\
           + i\lambda (\Ds\lambdab
                + [\sigma, \gamma_+\lambdab]
                + [\sigmab, \gamma_-\lambdab])
           + \Bigl(D + \frac{\sigma_2}{r}\Bigr)^2\Bigr],
\end{multline}
and the chiral multiplet action is
\begin{multline}
  \label{SC-5d}
  S_C = \frac{1}{2e^2} \int \! \sqrt{g} \, \rmd^5x \Tr\Bigl(
           \CFb{}^{\mu m} \CF_{\mu m}
           + \frac12 \cDb{}^m\sigmab\cD_m\sigma
           + \frac12 \cD^m\sigmab\cDb_m\sigma
           + 2iD D^m X_m \\
           - i\psib{}^m(\Ds\psi_m
                       - [\sigma, \gamma_-\psi_m]
                       - [\sigmab, \gamma_+\psi_m])
           + \psib{}^m \cD_m\lambda + \psi^m\cDb_m\lambdab
           + \Fb{}^m F_m\Bigr).
\end{multline}

It turns out that the superpotential of our theory is given by the
Chern-Simons functional
\begin{equation}
  W = \frac{1}{2} \int_M \ChS(\CA)
    = \frac12\int_M \Bigl(\CA\wedge\rmd\CA - \frac{2i}{3} \CA\wedge\CA\wedge\CA\Bigr).
\end{equation}
This choice gives
\begin{equation}
  \label{SW-5d}
  S_W = \frac{i}{2e^2} \int_{S^2} \! \sqrt{h} \, \rmd^2x \int_M
        \Tr\Bigl(F \wedge \CF - \frac{1}{2} \psi \wedge \rmd_\CA\psi
        + \Fb \wedge \CFb - \frac{1}{2} \psib \wedge \rmd_{\CAb}\psib\Bigr),
\end{equation}
where $\rmd_{\CA} = \rmd - i\CA$ and $\CF = \rmd\CA - i\CA \wedge \CA$
is the curvature of $\CA$.

To see that this is the right choice, take $M = \R^3$ and replace $h$
by the flat metric of $\R^2$, and drop from the action all the
curvature correction terms which depend explicitly on $r$.  One can
integrate out the auxiliary fields and show that the bosonic part of
the resulting expression can be written as
\begin{equation}
  \frac{1}{e^2} \int \! \rmd^5x
  \Tr\Bigl(\frac{1}{4} F^{MN} F_{MN}
    + \frac12 D^M X^I D_M X_I
    - \frac{1}{4} [X^I, X^J] [X_I, X_J]\Bigr),
\end{equation}
with $X^I = (\sigma_1, \sigma_2, X_3, X_4, X_5)$.  This is precisely
the bosonic part of the 5d MSYM action on $\R^5$.  (The same result is
obtained if $W$ is multiplied by an arbitrary phase factor, but such a
phase can be set to one by a $\U(1)_V$ rotation.)

There is, however, a problem with the above F-term.  In our theory the
superpotential $W$ has $q = 0$.  If the theory is formulated on $\R^2
\times M$, this simply means that $\U(1)_V$ is broken; $W$ must have
$q = 2$ for $S_W$ to have $q = 0$.  In fact, the lack of $\U(1)_V$ was
already apparent in our consideration of the twisting, where we saw
that the R-symmetry of the twisted theory includes $\U(1)_A$ but not
$\U(1)_V$.  On $S^2 \times M$, the consequence is more serious: $S_W$
is not supersymmetric unless $W$ has $q = 2$.

Still, one can find correction terms such that a half of $\CN = (2,2)$
supersymmetry on $S^2$ are preserved upon adding them to $S_W$.  The
supersymmetry variation of $S_W$ is
\begin{equation}
  \delta S_W =
  \frac{1}{2e^2 r} \int_{S^2}  \! \sqrt{h} \, \rmd^2x
  \bigl(-\psi_W \gamma_{\hat3}\veps + \psib_W \gamma_{\hat3}\vepsb\bigr),
\end{equation}
where $\psi_W$ is the fermion in the superpotential chiral multiplet
$(W, \psi_W, F_W)$.  For any $a \in \C^\times$, this can be
canceled by the supersymmetry variation of
\begin{equation}
  \label{SW'}
  S_W' = \frac{1}{2e^2 r} \int_{S^2}  \! \sqrt{h} \, \rmd^2x
         \bigl(aW - a^{-1} \Wb\bigr),
\end{equation}
provided that one imposes the parameters $(\veps, \vepsb)$ to satisfy
the relation
\begin{equation}
  \label{veps-vepsb}
  \veps = -a\gamma_{\hat 3} \vepsb.
\end{equation}
This relation halves the number of independent conformal Killing
spinors.  Hence, there is generally a family of F-terms $S_W + S_W'$
that preserve a half of the supersymmetries.

In the present case, we can actually determine the value of $a$ as
follows.  Suppose that $\CA$ is constant on the $S^2$.  Then $S_W'$ is
equal to
\begin{equation}
  \frac{2\pi r}{e^2} \bigl(a - a^{-1}\bigr) \Re W
  + \frac{2\pi ir}{e^2} \bigl(a + a^{-1}\bigr) \Im W.
\end{equation}
With $W$ proportional the Chern-Simons functional, for the path
integral to be well-defined with such terms the coefficient of $\Re W$
must be equal to an integer $k$, the level of the Chern-Simons
coupling, multiplied by a universal factor.  Since we want to keep $e$
and $r$ to be free parameters, the quantization of the level requires
$a = \pm 1$ and $k = 0$.  The two cases of opposite signs for $a$ are
related by orientation reversal of $M$, so we can take $a = -1$
without loss of generality.  Then we have
\begin{equation}
  \label{SW'-5d}
  S_W' = -\frac{1}{4e^2 r} \int_{S^2} \! \sqrt{h} \, \rmd^2x
         \int_M \bigl(\ChS(\CA) - \ChS(\CAb)\bigr),
\end{equation}
and $(\veps, \vepsb) = (\veps_\alpha, \vepsb_\alpha)$ satisfy the
constraint \eqref{veps-vepsb}.

To summarize, 5d MSYM can be formulated on $S^2 \times M$ preserving
the half of $\CN = (2,2)$ supersymmetry on $S^2$ generated by
\begin{equation}
  \label{p-supercharges}
  \Qb_\alpha + Q_\alpha.
\end{equation}
The total action is the sum
\begin{equation}
  S = S_V + S_C + S_W + S_W'
\end{equation}
of \eqref{SV-5d}, \eqref{SC-5d}, \eqref{SW-5d}, and \eqref{SW'-5d}.
We remark that the twisted theory is invariant under the rescaling
\begin{equation}
  \label{resc}
  e^2 \to \ell e^2, \quad
  r \to \ell r, \quad
  g_M \to \ell^2 g_M,
\end{equation}
as this can be absorbed by a scale transformation of the fields.

This is a good point to pause and discuss a possible alternative
formulation of the twisted theory, which we have ignored so far.  When
we determined the field content of the twisted theory as an $\CN =
(2,2)$ theory on $S^2$, after we worked out the transformation
properties of the fields we jumped to the conclusion that the theory
is described by a vector and chiral multiplets.  Actually, from the
transformation properties alone we cannot tell whether the fields come
from a vector and chiral multiplets, or from a twisted vector and
twisted chiral multiplets, since we do not know a priori whether
$\U(1)_R$ corresponds to $\U(1)_A$ or $\U(1)_V$.  $\CN = (2,2)$
twisted multiplets on $S^2$ are constructed in appendix \ref{TM}.

A twisted superpotential always preserves the full $\CN = (2,2)$
supersymmetry on $S^2$, provided that corrections similar to
\eqref{SW'} are included.  However, the twisted vector multiplet
action is problematic, being unbounded from below.  So if one wants to
formulate the theory using twisted multiplets, the best one can hope
for is to do that preserving some but not all of the supersymmetries.

The relevant supersymmetries are picked by a constraint imposed on the
parameters $(\veps, \vepsb)$.  The only constraints that are
compatible with the conformal Killing spinor conditions, and are
invariant under rotations of $S^2$, are those of the form
\eqref{veps-vepsb}, preserving the linear combinations $-a\Qb_\alpha +
Q_\alpha$.  Furthermore, $\U(1)_V$ must be broken to $\Z_2$, or it
would imply all the four supersymmetries (unless $a = 0$ or $\infty$,
but in these cases it is apparently impossible to write down a
sensible action).  Thus, as far as the symmetries are concerned, a
formulation using twisted multiplets can be as good as the formulation
using untwisted multiplets, but not better.  Besides, preserving two
supersymmetries with a twisted vector multiplet is hard, and we do not
know how to do this.


\section{TQFT in the infinite volume limit}
\label{TFT}

Having formulated the twisted theory on $S^2 \times M$, let us analyze
its dependence on the geometry of $M$.  We pick a supercharge $Q$
which we use as the BRST operator.  For definiteness we set
\begin{equation}
  \label{Q}
  Q =  \Qb_1 + Q_1,
\end{equation}
but the conclusions will be the same for any other linear combinations
of the preserved supercharges \eqref{p-supercharges}.  We will be
interested in quantities that are computed by path integrals only
involving $Q$-invariant operators and states.  Such quantities are
invariant under $Q$-exact shifts in the operators and states involved.

In many examples, twisting a supersymmetric theory leads to a TQFT.
What usually happens is that the action becomes independent of the
spacetime metric up to $Q$-exact terms, and therefore varying the
metric just inserts $Q$-exact operators in the path integral, leaving
$Q$-invariant quantities unchanged.  Since our theory is partially
twisted along $M$, one may expect that it is topological on $M$ modulo
$Q$-exactness.  This is not the \linebreak case, however.

In our theory the superpotential is topological, so the
$g_M$-dependence comes from $S_V$ and $S_C$.  These can be expressed
as
\begin{equation}
  \label{S-delta}
  \begin{aligned}
    S_V &= \bigl\{\Qb_1, [Q_2, V_V]\bigr\}
         = -\bigl\{Q_1, \bigl[\Qb_2, V_V\bigr]\bigr\}, \\
    S_C &= \bigl\{\Qb_1, [Q_2, V_C]\bigr\}
         = -\bigl\{Q_1, \bigl[\Qb_2, V_C\bigr]\bigr\},
  \end{aligned}
\end{equation}
with suitable functionals $V_V$ and $V_C$.%
\footnote{Explicitly,
  \begin{equation}
    \begin{aligned}
      V_V
      &= \frac{1}{2e^2}
         \int\!\sqrt{g}\, \rmd^5x\Tr\Bigl(
         \lambdab \gamma_{\hat3} \lambda - 4iD\sigma_2
         - \frac{2i}{r} \sigma_2^2\Bigr), \\
      V_C
      &= -\frac{1}{2e^2}
         \int\!\sqrt{g}\, \rmd^5x\Tr\Bigl(
         \psib^m \gamma_{\hat3} \psi_m - 4D^m X_m \sigma_2 +
         \frac{2i}{r} X^m X_m\Bigr).
    \end{aligned}
  \end{equation}
  To obtain $V_C$ from the corresponding formula in two dimensions,
  one uses the fact that for an adjoint chiral multiplet scalar $\phi$
  with $q = 0$, the equation
  \begin{equation}
    \CQb_\alpha \CQ_\beta \Tr \phib \phi
    = \frac{1}{2} \CQb_\alpha \CQ_\beta
      \Tr(2\phib \phi - \phi^2 - \phib^2)
    = 2 \CQb_\alpha \CQ_\beta \Tr(\Im\phi)^2
  \end{equation}
  holds modulo total derivatives.}
From these formulas we see that $S_V + S_C$ would be $Q$-exact if $Q$
were, say, $\Qb_1 + Q_2$ or $\Qb_2 + Q_1$.  However, it is not
$Q$-exact with respect to our choice \eqref{Q} or, for that purpose,
none of the preserved supercharges.  Thus the twisted theory depends
on $g_M$.

Nevertheless, one can argue that the twisted theory does become
topological on $M$ in the limit where the coupling $e^2$ and the
radius $r$ of the $S^2$ are rescaled as
\begin{equation}
  e^2 \to \ell^{-1} e^2, \quad
  r \to \ell^{-1} r
\end{equation}
and $\ell$ is sent to infinity, or equivalently by the scale
invariance \eqref{resc}, in the limit where $g_M$ is rescaled by an
infinitely large factor:
\begin{equation}
  \label{IVL}
  g_M \to \ell^2 g_M, \quad
  \ell \to \infty.
\end{equation}
Therefore, the infinite volume limit of the twisted theory is a TQFT.
We now present the argument.

Using the formulas \eqref{S-delta} we can write
\begin{equation}
  \label{S-Q-exact}
  \begin{aligned}
    S_V & = S_V^+ - \bigl\{Q, \bigl[\Qb_2, V_V\bigr]\bigr\}
          = S_V^- + \{Q, [Q_2, V_V]\}, \\
    S_C &= S_C^+ - \bigl\{Q, \bigl[\Qb_2, V_C\bigr]\bigr\}
         = S_C^- + \{Q, [Q_2, V_C]\},
  \end{aligned}
\end{equation}
where $S_V^+$ and $S_C^+$ have $q = +2$, while $S_V^-$ and $S_C^-$
have $q = -2$.  Thus a deformation of $g_M$ brings down operators
$T^\pm$ of $q = \pm 2$ into the correlation function:
\begin{equation}
  \frac{\delta}{\delta g^{mn}} \vev{\dotsb}
  = \vev{T^+_{mn} \dotsb}
  = \vev{T^-_{mn} \dotsb}.
\end{equation}

Now, the rescaling $g_M \to \ell^2 g_M$ is the same as keeping $g_M$
fixed but changing the \linebreak action to
\begin{equation}
  \label{S-ell}
  S_\ell = \ell^3 S_V + \ell S_C + S_W + S_W'.
\end{equation}
We see that in the limit $\ell \to \infty$, the F-term is negligible
compared to the other part of the action, hence $\U(1)_V$ is restored.
Then
\begin{equation}
  \label{Tt=0}
  \vev{T_{mn}^+ \dotsb} = 0,
\end{equation}
except possibly when the insertion $\dotsb$ has terms with $q = -2$.
This possibility is excluded by considering $\vev{T_{mn}^- \dotsb}$
instead, and we conclude that the theory is independent of $g_M$ in
the limit $\ell \to \infty$.

We were a bit sloppy when we asserted the equation \eqref{Tt=0} in the
above argument.  This point deserves a more careful explanation.

In the familiar story of compactification, one expands fields in the
eigenmodes of kinetic operators on the small ``internal'' space, and
their eigenvalues appear as masses from the point of view of the large
``external'' space.  Similarly, the nonzero modes of $\lambda$,
$\lambdab$ along the $S^2$ give masses of order $\ell^3$ from the
point of view of $M$, and those of $\psi_m$, $\psib_m$ give masses of
order $\ell$.  The kinetic terms describing the dynamics of these
modes on $M$ are accompanied by a factor of $\ell$.  Then natural
variables for the fermion integration for large $\ell$ are obtained by
rescaling
\begin{equation}
  \lambda_0 \to \ell^{-1/2} \lambda_0, \quad
  \lambda' \to \ell^{-3/2} \lambda', \quad
  \psi_m \to \ell^{-1/2}  \psi_m,
\end{equation}
where $\lambda_0$ and $\lambda'$ are the zero- and nonzero-mode parts
of $\lambda$, respectively.%
\footnote{For simplicity we assume that there are
no fermion zero modes on $M$.  This is generically true on an
odd-dimensional compact manifold since the relevant index is zero.}
The rescaling absorbs an infinite power of $\ell$ in the normalization
of the path integral measure.  On the other hand, the equations of
motion set
\begin{equation}
  F_m = -\frac{i}{2\ell} \eps_{mnp} \CFb{}^{np}.
\end{equation}
We write
\begin{equation}
  \lambda_0, \, \psi_m \sim \ell^{-1/2}, \quad
  \lambda' \sim \ell^{-3/2}, \quad
  F_m \sim \ell^{-1}
\end{equation}
to indicate the order of these fields according to the power of
$\ell$.

Since $S_V^+$ and $S_C^+$ have $q = 2$, they consist of terms with a
single $\Fb_m$ or two of $\lambdab$, $\psib_m$ (and terms with four or
more of these fields and fields with negative R-charge which we can
neglect).  Au such, they are at most of order $\ell^{-1}$:
\begin{equation}
  S^+_V  \lesssim \ell^{-1}, \qquad
  S^+_C  \lesssim \ell^{-1}.
\end{equation}
Then $T^+_{mn}$, derived from $\ell^3 S^+_V + \ell S^+_C$, is at
most of order $\ell^2$:
\begin{equation}
  T^+_{mn} \lesssim \ell^2.
\end{equation}
Meanwhile, $\U(1)_V$ is broken at order $\ell^{-1}$ by $S_W$.  All in
all, the counting suggests that $\vev{T^+_{mn} \dotsb}$ is at most
of order $\ell$,
\begin{equation}
  \vev{T^+_{mn} \dotsb} \lesssim \ell,
\end{equation}
and could grow like $\ell$ for large $\ell$, contrary to our previous
assertion.  Apparently our estimate was not very good.

So we need a more refined argument.  Let us consider a slightly
different setup, in which the action is taken to be
\begin{equation}
  \label{S_u}
  S_u = \ell^3 S_V + uS_C + S_W + S_W',
\end{equation}
with $\ell \leq u \leq \ell^3$.  First we show that the twisted theory
is independent of $u$ in the limit $\ell \to \infty$.

With the factor $l$ in front of $S_C$ replaced by $u$, this time we
have
\begin{equation}
  \lambda_0, \, \psi_m \sim u^{-1/2}, \quad
  \lambda' \sim \ell^{-3/2}, \quad
  F_m \sim u^{-1}.
\end{equation}
Following the same logic, we learn
\begin{equation}
  S^+_V \lesssim u^{-1}, \quad
  S^+_C \lesssim u^{-1},
\end{equation}
and $\U(1)_V$ is broken at order $u^{-1}$.  Then the change in the
correlation function between $u = \ell$ and $u = \ell'$ is
\begin{equation}
  \label{CFl1}
  \vev{\dotsb}_{u=\ell'} - \vev{\dotsb}_{u=\ell}
  = -\int_\ell^{\ell'} \! \rmd u \, \vev{S^+_C \dotsb}
  \lesssim \ell^{-1}.
\end{equation}
This goes to zero as $\ell \to \infty$, so the correlator is
independent of $u$ in this limit, as promised.

Using the freedom to choose $u$, we consider deformations of $g_M$ at
$u = \ell^3$.  Here we find that $T^+_{mn}$ is at most of order $1$
and
\begin{equation}
  \vev{T^+_{mn} \dotsb}_{u=\ell^3} \lesssim \ell^{-3}.
\end{equation}
This vanishes in the limit $\ell \to \infty$, which was what we wanted
to show.

Another way of understanding the topological invariance in the
infinite volume limit is to note that as a result of the restoration
of $\U(1)_V$, in this limit the full $\CN = (2,2)$ supersymmetry on
$S^2$ is restored and the $g_M$-dependent part of the action becomes
exact with respect to $\Qb_\alpha$ and $Q_\alpha$.  Then the
topological invariance of correlators is manifest for operator
insertions invariant under any of these supercharges, and this covers
a large class of $Q$-invariant operators.  This perspective also shows
clearly that these correlators are independent of $u$.

Moreover, choosing the BRST operator to be $\Qb_\alpha$ or $Q_\alpha$
allows to construct interesting observables.  For example, to a closed
loop $L \subset M$ one can associate a Wilson-loop-like operator
\begin{equation}
  \Tr P\exp\biggl(\int_{S^2} \! \sqrt{h}\, \rmd^2x \oint_M \CA\biggr),
\end{equation}
which is $\Qb_\alpha$-invariant.  So this is a very attractive
possibility, though justifying it surely requires a more careful
analysis.

\section{Localization to three dimensions}
\label{CS}

We have seen that the twisted theory on $S^2 \times M$ reduces to a
TQFT on $M$ in the limit where $M$ is infinitely large.  We now try to
determine this TQFT by exploiting one of the salient features of
supersymmetric theories: localization of the path integral.

Let us recall how the localization works.  Suppose one has a theory
with a fermionic charge $Q$, and wants to compute a correlation
function of $Q$-invariant operators.  To simplify the path integral,
one picks a $Q^2$-invariant functional $V$ such that the bosonic terms
in $\{Q, V\}$ are nonnegative.  Then one shifts the action as
\begin{equation}
  S \to S + t\{Q,V\}.
\end{equation}
By virtue of the $Q$-symmetry, the path integral is independent of
$t$.  Whereas one recovers at $t = 0$ the original path integral, the
path integral in the limit $t \to \infty$ localizes to the locus where
the bosonic terms in $\{Q, V\}$ all vanish.  In this limit the path
integral can be evaluated by first integrating over fluctuations
around a given configuration on the localization locus, and then over
all such background configurations.  In a typical situation the
localization locus is finite-dimensional.  Thus, the path integral
over an infinite-dimensional field space gets reduced to
finite-dimensional integrals.

For the purpose of computing physical quantities in our theory, this
localization method does not seem particularly useful; it is rather
difficult to find a good candidate for $V$ since $Q^2$ does not
generate a simple bosonic symmetry.  This is to be compared with the
situation where the action is $Q$-exact, in which case one can always
rescale the action by an infinitely large factor to go to the zero
coupling limit, whether $Q^2 = 0$ or not.  By contrast, none of the
$Q$-invariant pieces of our action is $Q$-exact.  One might be tempted
to rescale the $Q$-exact terms in the expressions \eqref{S-Q-exact},
but doing so would break the $Q$-symmetry as these terms are not
$Q$-invariant by themselves.

However, we are interested in the infinite volume limit of $M$, and
the things are better here.  As we saw in the previous section, in
this limit we can replace the action by the one given in \eqref{S_u},
so let us define $t = \ell^3$ and set $u = t$:
\begin{equation}
  S_t = t(S_V + S_C) + S_W + S_W'.
\end{equation}
This is just like what one would do to localize the path integral if
$S_V + S_C$ were $Q$-exact, namely, rescale
\begin{equation}
  S_V + S_C \to t(S_V + S_C)
\end{equation}
and take $t \to \infty$.  Then localization should take place and it
should be possible to rewrite the $t \to \infty$ limit of the path
integral with respect to $S_t$ as an integral over a smaller space.
Hopefully, the expression after the localization can be interpreted as
the path integral for some three-dimensional TQFT.

Let us analyze the localization property of this path integral.
Integrating out the auxiliary fields $F_m$, $\Fb_m$ leaves the
potential
\begin{equation}
  \label{F-term}
  \frac{1}{4e^2 t} \int \! \sqrt{g} \, \rmd^5x
  \Tr \CFb{}^{mn}\CF_{mn},
\end{equation}
which disappears in the limit $t \to \infty$.  The rest of the
bosonic terms in $S_V$ and $S_C$ are nonnegative (except one imaginary
term) and multiplied by a factor of $t$.  Thus the path integral
localizes to the saddle-point configurations:
\begin{equation}
  0 = F_{\hat1\hat2} + \frac{\sigma_1}{r}
    = D_\mu\sigma_1
    = D_\mu\sigma_2
    = [\sigma_1, \sigma_2]
    = \CF_{\mu m}
    = \cD_m\sigma_1
    = \cD_m\sigma_2
    = D^m X_m.
\end{equation}

For these configurations, the gauge field $A_{S^2} = A_\mu \rmd x^\mu$
on $S^2$ solves the Yang-Mills equation $D^\mu F_{\mu\nu} = 0$.  In
two dimensions gauge fields have no physical propagating degrees of
freedom, so in each topological class there is a unique solution up to
gauge transformations.  Convenient representatives are
\begin{equation}
  A_{S^2}^\pm = B(\pm 1 - \cos\theta) \rmd\varphi,
\end{equation}
where $\pm$ refers to the chart covering the north or south pole, and
\begin{equation}
  B = \frac{1}{2\pi} \int_{S^2} F_{\hat1\hat2}
\end{equation}
is the magnetic flux through the $S^2$ which can be chosen to be
diagonal.  Each diagonal entry of $B$ represents the Chern number of a
line bundle and must be quantized.  The residual gauge symmetry is the
group of three-dimensional gauge transformations
\begin{equation}
  g\colon M \to G_B
\end{equation}
valued in the stabilizer $G_B \subset G$ of $B$.

For the above choice of $A_{S^2}$, we have
\begin{equation}
  \sigma_1 = \frac{B}{2r}.
\end{equation}
Then $D_\mu\sigma_1 = 0$ is automatically satisfied.  On the other
hand, $\cD_m\sigma_1 = 0$ shows that $\CA$ takes values in $\gf_B
\otimes \C$, with $\gf_B$ the Lie algebra of $G_B$.  Given this,
$\CF_{\mu m} = 0$ implies that $\CA$ is constant on the $S^2$:
\begin{equation}
  \CA_m = \CA_m(x^m) \in \gf_B \otimes \C.
\end{equation}
The imaginary part of $\CA$ obeys the D-term equation $\mu(\CA) = 0$,
where
\begin{equation}
  \label{mu}
  \mu(\CA) = D^m X_m.
\end{equation}
Likewise, $D_\mu\sigma_2 = [\sigma_1, \sigma_2] = 0$ implies that
$\sigma_2$ is constant on the $S^2$ and valued in $\gf_B$,
\begin{equation}
  \sigma_2 = \sigma_2(x^m) \in \gf_B,
\end{equation}
and the last equation
\begin{equation}
  \cD_m \sigma_2 = 0
\end{equation}
says $\sigma_2$ is covariantly constant on $M$.

Now that we know the saddle-point configurations, we can write down
the localized expression of the path integral.  If we denote by $(B,
\CA_B, a_B)$ the saddle-point configuration with a magnetic flux $B$
and $(\CA, \sigma_2) = (\CA_B, a_B)$, then the partition function is
\begin{equation}
  \sum_B \int \cD\CA_B \, \rmd a_B \, Z_1(B, \CA_B, a_B)
  \exp\bigl(-S_0(\CA_B)\bigr).
\end{equation}
Here $Z_1(B, \CA_B, a_B)$ represents the one-loop determinants for
fluctuations around the background $(B, \CA_B, a_B)$, and
\begin{equation}
  S_0(\CA_B) = -\frac{\pi r}{e^2} \int_M \bigl(\ChS(\CA_B) - \ChS(\CAb_B)\bigr)
\end{equation}
is the action evaluated at the background.  Correlation functions of
$Q$-invariant operators are given by inserting the operators evaluated
at the saddle-point configurations in the above path integral.

Suppose we have done the integrations over the $a_B$.  Then, for each
$B$, we are left with the path integral
\begin{equation}
  \label{LPI'}
  \int \widetilde{\cD\CA_B} \exp(-S_0(\CA_B)),
\end{equation}
with the modified measure
\begin{equation}
  \widetilde{\cD\CA_B} = \cD\CA_B \int \rmd a_B \,  Z_1(B, \CA_B, a_B).
\end{equation}
Written in this form, it is clear that each summand of the localized
path integral is the path integral for a three-dimensional gauge
theory, whose action is independent of $g_M$.  This is in keeping with
the expectation that the twisted theory reduces to a three-dimensional
TQFT in the infinite volume limit.

There is, however, something not very beautiful about the expression
\eqref{LPI'}.  The field variable of the path integral is a gauge
field $\CA_B$ for the complex gauge group $(G_B)_\C$.  Moreover, the
action $S_0(\CA_B)$ is invariant under the complex gauge
transformations.  These facts strongly suggest that we should
interpret it as the path integral for a gauge theory with gauge group
$(G_B)_\C$.  And yet, the integration is performed within the space of
complex gauge fields modulo \emph{real} gauge transformations, since
the gauge group of the original theory is $G$, not $G_\C$.  The
question is then whether we can convert the gauge group from \linebreak $G_B$ to
$(G_B)_\C$.

To phrase this more precisely, let $\CU$ be the space of real gauge
fields and $\CG$ the group of real gauge transformations, and let
$\CU_\C$ and $\CG_\C$ be the complexifications of $\CU$ and $\CG$.
(The discussion will be identical for all values of $B$, so we will
not specify it to avoid cluttering the notation.)  Then the
integration domain is
\begin{equation}
  \mu^{-1}(0)/\CG \subset \CU_\C/\CG,
\end{equation}
where $\mu$ was defined in \eqref{mu}.  The question is, can we embed
this space into the moduli space $\CU_\C/\CG_\C$ of complex gauge
fields, rather than $\CU_\C/\CG$?

The answer is yes, and indeed it is done naturally.  To understand
how, recall a standard fact in supersymmetric gauge theories about the
moduli space of vacua.  This is the space of solutions to the D- and
F-term equations (the equations obtained by setting to zero the D- and
F-term potentials) modulo gauge transformations.  Besides this
description, there is another way to describe the same space: one
drops the D-term equation and takes the quotient instead by the
complexified gauge transformations.

In our situation, the D-term equation is $\mu = 0$, while the F-term
equation is absent as the potential \eqref{F-term} becomes zero in the
limit $t \to \infty$.  Then the two descriptions of the moduli space
lead to the identification of $\mu^{-1}(0)/\CG$ and $\CU_\C/\CG_\C$.
In fact, relevant points in $\CU_\C$ are only those that can be mapped
into the locus $\mu = 0$ by a complex gauge transformation; such
points are called semistable.  Thus we have
\begin{equation}
   \mu^{-1}(0)/\CG \iso \CU_\C^{\text{ss}}/\CG_\C,
\end{equation}
where $\CU_\C^{\text{ss}}$ is the set of semistable points of
$\CU_\C$. Mathematically, $\mu$ is (the dual of) the moment map for
the $\CG$-action on $\CU_\C$ with respect to the K\"ahler form
\begin{equation}
  \omega
  = \frac{i}{2} \int_M \Tr\bigl(\delta \CA \wedge \star \delta \CAb\bigr),
\end{equation}
and we have an identification of the symplectic and GIT quotients.

Therefore, we arrive at the expression
\begin{equation}
  \label{LPI}
  \sum_B \int_{\CM_B^{\text{ss}}}
  \widetilde{\cD\CA_B} \exp\bigl(-S_0(\CA_B)\bigr),
\end{equation}
where $\CM_B^{\text{ss}}$ is the semistable locus in the moduli space
$\CM_B$ of $(G_B)_\C$ gauge fields.  

Apart from the modifications in the measure and the integration
domain, each summand of the above expression is the path integral for
Chern-Simons theory on $M$ with gauge group $(G_B)_\C$:
\begin{equation}
  \int_{\CM_B} \cD\CA_B \exp\bigl(iS_{\ChS}(\CA_B)\bigr).
\end{equation}
The complex Chern-Simons action is given by
\begin{equation}
  S_\ChS
  = -\frac{k + is}{8\pi} \int_M \ChS(\CA)
    - \frac{k - is}{8\pi} \int_M \ChS(\CAb),
\end{equation}
with $k \in \Z$ and $s \in \R$.  The comparison shows
\begin{equation}
  k = 0, \quad
  s = \frac{8\pi^2 r}{e^2}.
\end{equation}
Notice that the parameter $s$ is invariant under a rescaling of $e^2$
and $r$ by a common factor.  This is consistent with the fact that the
twisted theory is invariant under the rescaling~\eqref{resc} and
becomes topological on $M$ in the infinite volume limit.

\section*{Acknowledgments}

I would like to thank Sungjay Lee and Masahito Yamazaki for
discussions.  This work is supported by National University of
Singapore Start-up Grant R144-000-269-133.

\appendix

\section{Spinors on \texorpdfstring{\boldmath{$S^2$}}{S**2}}
\label{S2}

The metric of a round two-sphere $S^2$ of radius $r$ is
\begin{equation}
   h = r^2 (\rmd\theta^2 + \sin^2\!\theta \, \rmd\varphi^2),
\end{equation}
with $(\theta, \varphi)$ the spherical coordinates.  We set
\begin{equation}
  (x^1, x^2) = (\theta, \varphi),
\end{equation}
and introduce the orthonormal vectors
\begin{equation}
  e_{\hat1} = \frac{1}{r} \frac{\del}{\del\theta}, \qquad
  e_{\hat2} = \frac{1}{r\sin\theta} \frac{\del}{\del\varphi}.
\end{equation}
We use the indices $\mu$, $\nu$, $\dotsc$ to refer to the coordinates,
and $\muh$, $\nuh$, $\dotsc$ to refer to the orthonormal frame.

We choose the gamma matrices representing the Clifford algebra
$\{\gamma_{\muh}, \gamma_{\nuh}\} = 2\delta_{\hat\mu\hat\nu}$ to be
\begin{equation}
  \gamma_{\muh} = \tau_{\hat\mu},
\end{equation}
where $\tau_{\hat\mu}$ are the Pauli matrices:
\begin{equation}
  \tau_{\hat1}
  = \biggl(\begin{array}{cc}
      0 & 1 \\
      1 & 0
    \end{array}\biggr), \qquad
  \tau_{\hat2}
  = \biggl(\begin{array}{cc}
      0 & -i \\
      i & 0
    \end{array}\biggr), \qquad
  \tau_{\hat3}
  = \biggl(\begin{array}{cc}
      1 & 0 \\
      0 & -1
    \end{array}\biggr).
\end{equation}
The chirality operator is
\begin{equation}
  \gamma_{\hat3} = \tau_{\hat3}.
\end{equation}
The spin connection is denoted by $\nabla$.

The product of two spinors $\psi$ and $\chi$ is defined as
\begin{equation}
  \psi\chi = \psi^T C \chi,
\end{equation}
where the charge conjugation operator
\begin{equation}
  C = i\tau_{\hat2}.
\end{equation}
For anticommuting spinors, we have
\begin{equation}
  \psi\chi = \chi\psi, \qquad
  \psi\gamma_{\ah}\chi = -\chi\gamma_{\ah}\psi \qquad
  (\ah = \hat1, \, \hat2, \, \hat3).
\end{equation}
More generally,
\begin{equation}
  \psi\gamma_{\ah_1}\dotsb\gamma_{\ah_n}\chi
  = (-1)^{n(n+1)/2} \chi\gamma_{\ah_1}\dotsb\gamma_{\ah_n}\psi
\end{equation}
for the $\ah_i$ all different.

The following identities are useful in calculating supersymmetry variations:
\begin{align}
  2(\veps\eta)(\alpha\beta)
  &= (\veps\alpha)(\eta\beta)
    - \sum_{\ah = \hat1}^{\hat3} (\veps\gamma_{\ah}\alpha)(\eta\gamma_{\ah}\beta), \\
  (\veps\eta)(\alpha\beta) - (\veps\gamma_{\hat3}\eta)(\alpha\gamma_{\hat3}\beta)
  &= (\veps\alpha)(\eta\beta) - (\veps\gamma_{\hat3}\alpha)(\eta\gamma_{\hat3}\beta).
\end{align}
Here $\veps$, $\eta$ are commuting spinors, and $\alpha$, $\beta$ are
anticommuting spinors.

\section{\texorpdfstring{\boldmath{$\CN = (2,2)$}}{N = (2,2)}
  supersymmetry on \texorpdfstring{\boldmath{$S^2$}}{S**2}}
\label{SUSY}

In this appendix we review the construction of $\CN = (2,2)$
supersymmetric gauge theories on $S^2$.  We refer the reader to
\cite{Benini:2012ui, Doroud:2012xw} for more discussions.

The $\CN = (2,2)$ supersymmetry group on $S^2$ is a subgroup of the
$\CN = (2,2)$ superconformal group on $S^2$ that does not include
conformal transformations.  The latter group is generated by the
rotations and conformal transformations of $S^2$, the vector
R-symmetry $\U(1)_V$, and the axial R-symmetry $\U(1)_A$, as well as
eight supersymmetries.  On fermionic fields $\Psi$ which are Dirac
spinors, the R-symmetries act by
\begin{equation}
  \begin{aligned}
    \U(1)_V \ni e^{i\alpha} &\colon \Psi \mapsto \exp(iq_V\alpha) \Psi, \\
    \U(1)_A \ni e^{i\beta} &\colon \Psi \mapsto \exp(iq_A\beta\gamma_{\hat3}) \Psi,
  \end{aligned}
\end{equation}
where $q_V$ and $q_A$ are the vector and axial R-charges of $\Psi$.
On bosonic fields they both act in the same way, without the
$\gamma_{\hat3}$ insertion for $\U(1)_A$.

There are two kinds of supercharges, $\Qb_\veps$ and $Q_{\vepsb}$,
labeled by commuting conformal Killing spinors $\veps$ or $\vepsb$
which can be chosen to obey
\begin{equation}
  \nabla_\mu\veps = \pm\frac{1}{2r}\gamma_\mu\gamma_{\hat3}\veps, \quad
  \nabla_\mu\vepsb = \pm\frac{1}{2r}\gamma_\mu\gamma_{\hat3}\vepsb.
\end{equation}
They satisfy the commutation relations
\begin{equation}
  \label{SUSY-1}
  \{\Qb_{\veps_1}, \Qb_{\veps_2}\} =
  \{Q_{\vepsb_1}, Q_{\vepsb_2}\} = 0
\end{equation}
and
\begin{equation}
  \label{SUSY-2}
  \{\Qb_{\veps}, Q_{\vepsb}\}
  = P_\xi + i\alpha F_V + i\beta F_A.
\end{equation}
Here $P_\xi$ generates the rotations and conformal transformations by
the conformal Killing vector $\xi$, and $F_V$ and $F_A$ are the
generators of $\U(1)_V$ and $\U(1)_A$, respectively; the parameters
are given by
\begin{equation}
  \xi^\mu = i\veps\gamma^\mu\vepsb, \qquad
  \alpha = -\frac{1}{4}(\veps\nablas\vepsb - \vepsb\nablas\veps), \qquad
  \beta = \frac{1}{4}(\veps\gamma_{\hat3}\nablas\vepsb - \vepsb\gamma_{\hat3}\nablas\veps).
\end{equation}

Two important types of field representations of the $\CN = (2,2)$
superconformal group are a vector multiplet
\begin{equation}
  (A, \sigma, \lambda, \lambdab, D)
\end{equation}
consisting of a gauge field $A$, a complex scalar $\sigma$, Dirac
spinors $\lambda$, $\lambdab$, and a real auxiliary field $D$, and
a chiral multiplet
\begin{equation}
  (\phi, \psi, F)
\end{equation}
consisting of a complex scalar $\phi$, a Dirac spinor $\psi$, and a
complex auxiliary field $F$.  The Weyl weight $w$ and the vector
R-charge $q$ of a chiral multiplet are related by
\begin{equation}
  q = 2w.
\end{equation}
The following table summarizes the Weyl weight, vector R-charge, and axial
R-charge of the component fields:
\begin{equation}
\renewcommand{\arraystretch}{1.2}
  \begin{array}{|@{\hskip .5em}c@{\hskip .5em}|
      @{\hskip .5em}c@{\hskip 1em}c@{\hskip 1em}c
      @{\hskip .7em}c@{\hskip .7em}c@{\hskip .5em}|
      @{\hskip .5em}c@{\hskip 1em}c@{\hskip 1em}c@{\hskip .5em}|}
    \hline
   & A_\mu & \sigma  & \lambda & \lambdab & D &
    \phi & \psi & F \\ \hline
    w  & 0 & 1  & 3/2 & 3/2 & 2
       & q/2 & (q + 1)/2 &  (q + 2)/2 \\
   F_V & 0 & 0 & -1 & 1 & 0
       & q & q - 1 & q - 2 \\
   F_A & 0 & 2 & -1 & 1 & 0
       & 0 & 1 & 0 \\
       \hline
  \end{array}
\end{equation}

The supersymmetry transformation of a vector multiplet is
\begin{equation}
  \begin{aligned}
   \delta A_\mu &= -\frac{i}{2} (\vepsb\gamma_\mu\lambda
                                + \veps\gamma_\mu\lambdab), \\
   \delta\sigma &= \vepsb\gamma_-\lambda - \veps\gamma_+\lambdab, \\
   \delta\sigmab &= \vepsb\gamma_+\lambda - \veps\gamma_-\lambdab, \\
   \delta\lambda
   &= i\Bigl(F_{\hat1\hat2} \gamma_{\hat3}
            + \gamma_-\Ds\sigma
            + \gamma_+\Ds\sigmab
            + \frac{1}{2} [\sigma, \sigmab] \gamma_{\hat3} + iD\Bigr)\veps
            + i(\sigma\gamma_- + \sigmab\gamma_+) \nablas\veps, \\
   \delta\lambdab
   &= i\Bigl(F_{\hat1\hat2} \gamma_{\hat3}
             - \gamma_+\Ds\sigma
             - \gamma_-\Ds\sigmab
            - \frac{1}{2} [\sigma, \sigmab] \gamma_{\hat3} - iD\Bigr)\vepsb
            - i(\sigma\gamma_+ + \sigmab\gamma_-) \nablas\vepsb, \\
   \delta D &= -\frac{i}{2}\vepsb
               (\Ds\lambda
                + [\sigma, \gamma_+\lambda]
                + [\sigmab, \gamma_-\lambda])
               - \frac{i}{2} \lambda\nablas\vepsb \\
               &\quad
               + \frac{i}{2}\veps
                  (\Ds\lambdab
                  - [\sigma, \gamma_-\lambdab]
                  - [\sigmab, \gamma_+\lambdab])
               + \frac{i}{2} \lambdab\nablas\veps,
\end{aligned}
\end{equation}
where $\gamma_\pm = (1 \pm \gamma_{\hat3})/2$ are the projectors to the
positive- and negative-chirality subspaces.  The supersymmetry
transformation of a chiral multiplet coupled to a vector multiplet is
\begin{equation}
  \begin{aligned}
    \delta\phi &= \vepsb\psi, \\
    \delta\phib &= \veps\psib, \\
    \delta\psi &= i(\Ds\phi + \sigma\phi\gamma_+ + \sigmab\phi\gamma_-)\veps
                  + i\frac{q}{2} \phi\nablas\veps + F\vepsb, \\
    \delta\psib &= i(\Ds\phib + \phib\sigma\gamma_- + \phib\sigmab\gamma_+)\vepsb
                  + i\frac{q}{2} \phib\nablas\vepsb + \Fb\veps, \\
   \delta F &=  i\veps(\Ds\psi - \gamma_-\sigma\psi - \gamma_+\sigmab\psi
                   - \lambda\phi) + i\frac{q}{2} \psi\nablas\veps, \\
   \delta\Fb &= i\vepsb(\Ds\psib - \gamma_+\psib\sigma - \gamma_-\psib\sigmab
                       + \phib\lambdab) + i\frac{q}{2} \psib\nablas\vepsb.
  \end{aligned}
\end{equation}
These transformation rules satisfy the commutation relations
\eqref{SUSY-1}, \eqref{SUSY-2} modulo gauge transformations; see
\cite{Benini:2012ui} for details of the calculation.

On $S^2$ there are four independent conformal Killing spinors, leading
to the total of eight supercharges.  The supercharges of $\CN = (2,2)$
supersymmetry are obtained by restricting $\veps$, $\vepsb$ to satisfy
\begin{align}
  \label{CKS-1}
  \nabla_\mu\veps &= +\frac{1}{2r}\gamma_\mu\gamma_{\hat3}\veps, \\
  \label{CKS-2}
  \nabla_\mu\vepsb &= -\frac{1}{2r}\gamma_\mu\gamma_{\hat3}\vepsb.
\end{align}
For this choice of the signs, $\xi$ is a Killing vector and generates
an isometry, and $\beta = 0$.  Thus the $\CN = (2,2)$ supersymmetry
group is generated by rotations, $\U(1)_V$, and four supersymmetry
transformations.

We pick conformal Killing spinors $\veps_\alpha$ ($\alpha = 1$, $2$)
that satisfy \eqref{CKS-1} and
\begin{equation}
  \veps_1\veps_2 = 1.
\end{equation}
Concretely, we can choose
\begin{equation}
  \veps_1
  = e^{i \varphi/2}
  \biggl(\begin{array}{c}
    \cos(\theta/2) \\ \sin(\theta/2)
  \end{array}\biggr), \qquad
  \veps_2
  = e^{-i \varphi/2}
  \biggl(\begin{array}{c}
    -\sin(\theta/2) \\ \cos(\theta/2)
  \end{array}\biggr).
\end{equation}
We set $\vepsb_\alpha = \gamma_{\hat3}\veps_\alpha$.  Then $\vepsb_\alpha$
satisfy \eqref{CKS-2} and $\vepsb_1\vepsb_2 = -1$.  We write
$\Qb_{\alpha}$ for $\Qb_{\veps_\alpha}$ and $Q_\alpha$ for
$Q_{\vepsb_\alpha}$.  The corresponding supersymmetry variations are
denoted by $\CQb_\alpha$, $\CQ_\alpha$.

The vector multiplet action $S_V$ and the chiral multiplet action
$S_C$ are given by
\begin{align}
  (\veps_\alpha\veps_\beta) S_V
  &= \CQb_\alpha\CQ_\beta \int\!\sqrt{h}\, \rmd^2x\Tr\biggl(
         \lambdab \gamma_{\hat3} \lambda - 4iD\sigma_2
         - \frac{2i}{r} \sigma_2^2\biggr), \\
  (\veps_\alpha\veps_\beta) S_C
  &= -\CQb_\alpha\CQ_\beta \int\!\sqrt{h}\, \rmd^2x\Tr\biggl[
           \psib \gamma_{\hat3} \psi
           - 2\phib\Bigl(\sigma_2 + i\frac{q}{2r}\Bigr)\phi
           + \frac{i}{r} \phib\phi\biggr].
\end{align}
These formulas together with the relations $\CQb_\alpha^2 =
\CQ_\alpha^2 = 0$ show that $S_V$ and $S_C$ are invariant under $\CN =
(2,2)$ supersymmetry.  More explicitly,
\begin{align}
  \label{S_V}
   \begin{split}
    S_V &= \int \! \sqrt{h} \, \rmd^2x \Tr \biggl[ \Bigl(F_{\hat1\hat2}
    + \frac{\sigma_1}{r}\Bigr)^2 + D^\mu\sigmab D_\mu\sigma
    + \frac{1}{4} ([\sigma, \sigmab])^2
    \\ &\quad
    + i\lambda (\Ds\lambda + [\sigma, \gamma_+\lambda]
        + [\sigmab,  \gamma_-\lambda])
    + \Bigl(D + \frac{\sigma_2}{r}\Bigr)^2\biggr],
  \end{split} \\
  \label{S_C}
  \begin{split}
    S_C &= \int \! \sqrt{h} \, \rmd^2x \biggl[\phib\biggl(-D_\mu^2 +
    \frac{1}{2}\{\sigma, \sigmab\} + iD + i\frac{q}{r} \sigma_2
    - \frac{q^2 - 2q}{4r^2}\biggr)\phi
    \\ &\quad
    - i\psib\Bigl(\Ds - \sigma\gamma_- - \sigmab\gamma_+ +
    \frac{q}{2r} \gamma_{\hat3}\Bigr)\psi + i\psib\lambda\phi -
    i\phib\lambdab\psi
    + \Fb F\biggr],
  \end{split}
\end{align}
where $\sigma = \sigma_1 - i\sigma_2$.

Given a gauge-invariant holomorphic function $W(\phi)$ of chiral
multiplet scalars $\phi^i$ with total vector R-charge $q = 2$, one can
construct an F-term $S_W$ that is invariant under $\CN = (2,2)$
supersymmetry.  The superpotential $W$ may itself be thought of as the
lowest component of a chiral multiplet $(W, \psi_W, F_W)$, with the
auxiliary field
\begin{equation}
  F_W = F^i \del_i W
        - \frac{1}{2} \psi^i\psi^j \del_i\del_j W.
\end{equation}
In terms of this multiplet,
\begin{equation}
  \label{SP}
  S_W = i\int \! \sqrt{h} \, \rmd^2x (F_W + \Fb_W).
\end{equation}
Note that $S_W$ must be purely imaginary in order to produce a
positive F-term potential.

\section{\texorpdfstring{\boldmath{$\CN =  (2,2)$}}{N = (2,2)} twisted
  multiplets on \texorpdfstring{\boldmath{$S^2$}}{S**2}}
\label{TM}

$\CN = (2,2)$ vector and chiral multiplets have ``twisted'' cousins of
the identical field content but with different R-charge assignments.
A twisted vector multiplet
\begin{equation}
  (A, \rho, \kappa, \kappab, C)
\end{equation}
consists of a gauge field $A$, a complex scalar $\rho$, Dirac spinors
$\kappa$, $\kappab$, and a real auxiliary field $C$; a twisted chiral
(plus antichiral) multiplet
\begin{equation}
  (\ups, \chi, \chib, E)
\end{equation}
consists of a complex scalar $\ups$, Dirac spinors $\chi$, $\chib$, and
a complex auxiliary field $E$.  Their Weyl weights and R-charges are
as follows:
\begin{equation}
\renewcommand{\arraystretch}{1.2}
  \begin{array}{|@{\hskip .5em}c@{\hskip .5em}|
      @{\hskip .5em}c@{\hskip 1em}c@{\hskip 1em}c
      @{\hskip .7em}c@{\hskip .7em}c@{\hskip .5em}|
      @{\hskip .5em}c@{\hskip 1em}c@{\hskip 1em}c@{\hskip .5em}|}
    \hline
   & A_\mu & \rho & \kappa & \kappab & C &
    \ups & \chi & E \\ \hline
    w  & 0 & 1  & 3/2 & 3/2 & 2
       & q/2 & (q + 1)/2 &  (q + 2)/2 \\
   F_V & 0 & -2 & -1 & 1 & 0
       & 0 & -1 & 0 \\
   F_A & 0 & 0 & -1 & 1 & 0
       & -q & 1 - q & 2 - q \\
\hline
  \end{array}
\end{equation}

The supersymmetry transformation rules for twisted multiplets can be
obtained from those for the corresponding untwisted multiplets by
replacing
\begin{equation}
  \label{map-eps}
  \begin{aligned}
    \veps  &\to \gamma_- \veps + \gamma_+ \vepsb, \\
    \vepsb &\to \gamma_- \vepsb + \gamma_+ \veps,
  \end{aligned}
\end{equation}
and mapping the fields properly.  The closure of the superconformal
algebra is ensured by the fact that the right-hand sides are again
conformal Killing spinors.  In the case of the flat-space
supersymmetry, the parameters can be chosen to be constant spinors of
definite chirality, $\veps_\pm$ and $\vepsb_\pm$ with $\gamma_{\hat3}
= \pm 1$, and the replacement is merely the relabeling of the
parameters $\veps_+ \to \vepsb_+$, \ $\vepsb_+ \to \veps_+$; the
distinction between twisted and untwisted multiplets is therefore a
matter of convention.  On $S^2$, this procedure gives essentially
different multiplets since one cannot choose the conformal Killing
spinors $\veps_\alpha$ (and hence also $\vepsb_\alpha =
\gamma_{\hat3}\veps_\alpha$) to have definite chirality.

The replacement \eqref{map-eps} has the effect of exchanging $\alpha
\leftrightarrow -\beta$ in the commutation relation \eqref{SUSY-2}.
This is harmless for bosonic fields as $\U(1)_V$ and $\U(1)_A$ act in
the same way on them, so this just switches the charges.  However, to
maintain the supersymmetry algebra on the fermions, we must map them
in a similar fashion to \eqref{map-eps} so that $\U(1)_V$ and
$\U(1)_A$ are exchanged.  We map the vector to twisted vector
multiplet fields as
\begin{equation}
  \begin{aligned}
    \sigma &\to \rho, \\
    \lambda  &\to \gamma_- \kappa + \gamma_+ \kappab, \\
    \lambdab &\to \gamma_- \kappab + \gamma_+ \kappa, \\
    D &\to C,
  \end{aligned}
\end{equation}
and the chiral to twisted chiral multiplet fields as
\begin{equation}
  \begin{aligned}
    \phi &\to \ups, \\
    \psi  &\to \gamma_- \chib + \gamma_+ \chi, \\
    \psib &\to \gamma_- \chi + \gamma_+ \chib, \\
    F &\to E.
  \end{aligned}
\end{equation}
The latter mixes the representation $R$ of the multiplets with its
complex conjugate $\Rb$, so makes sense only if $R$ is real;
implicitly assumed is an isomorphism that exchanges $R \leftrightarrow
\Rb$ and commutes with gauge transformations.

Applying the above prescription, we find that the supersymmetry
transformation of a twisted vector multiplet is
\begin{equation}
  \begin{aligned}
   \delta A_\mu &= -\frac{i}{2} (\vepsb\gamma_\mu\kappa
                                + \veps\gamma_\mu\kappab), \\
   \delta\rho &= -\veps\gamma_{\hat3}\kappa, \\
   \delta\rhob &= \vepsb\gamma_{\hat3}\kappab, \\
   \delta\kappa
   &= i\Bigl(F_{\hat1\hat2}\gamma_{\hat3} - \frac{1}{2} [\rho, \rhob] - iC\gamma_{\hat3}\Bigr)\veps
      - i\gamma_{\hat3}\Ds\rho \vepsb
      - i\rho\gamma_{\hat3} \nablas\vepsb, \\
   \delta\kappab
   &= i\Bigl(F_{\hat1\hat2}\gamma_{\hat3} + \frac{1}{2} [\rho, \rhob] + iC\gamma_{\hat3}\Bigr)\vepsb
      + i\gamma_{\hat3}\Ds\rhob \veps
      + i\rhob\gamma_{\hat3} \nablas\veps, \\
   \delta C &= \frac{i}{2}\veps
               (\gamma_{\hat3}\Ds\kappab - [\rhob, \kappa])
               - \frac{i}{2} \kappab\gamma_{\hat3}\nablas\veps
               -\frac{i}{2}\vepsb
                (\gamma_{\hat3}\Ds\kappa + [\rho, \kappab])
               + \frac{i}{2} \kappa\gamma_{\hat3}\nablas\vepsb,
\end{aligned}
\end{equation}
and that of a twisted chiral multiplet is
\begin{equation}
  \label{TCM-SUSY}
  \begin{aligned}
    \delta\ups &= \vepsb\gamma_+\chi + \veps\gamma_-\chib, \\
    \delta\upsb &= \veps\gamma_+\chib + \vepsb\gamma_-\chi, \\
    \delta\chi
    &= \gamma_+(i\Ds\ups\veps + i\rho\ups\vepsb + E\veps)
       + \gamma_-(i\Ds\upsb\veps - i\rho\upsb\vepsb + \Eb\veps)
       + i\frac{q}{2} (\ups\gamma_+ + \upsb\gamma_-)\nablas\veps, \\
    \delta\chib
    &= \gamma_- (i\Ds\ups\vepsb + i\rhob\ups\veps + E\vepsb)
       + \gamma_+ (i\Ds\upsb\vepsb - i\rhob\upsb\veps + \Eb\vepsb)
       + i\frac{q}{2} (\ups\gamma_- + \upsb\gamma_+)\nablas\vepsb, \\
    \delta E &= i\veps\gamma_+(\Ds\chib - \rhob\chi  - \kappab\ups)
                + i\vepsb\gamma_-(\Ds\chi - \rho\chib - \kappa\ups)
                + i\frac{q}{2} (\chib\gamma_+\nablas\veps
                + \chi\gamma_-\nablas\vepsb), \\
    \delta\Eb &= i\vepsb\gamma_+(\Ds\chi + \rho\chib - \kappa\upsb)
                + i\veps\gamma_-(\Ds\chib + \rhob\chi - \kappab\upsb)
                + i\frac{q}{2} (\chi\gamma_+\nablas\vepsb
                + \chib\gamma_-\nablas\veps).
  \end{aligned}
\end{equation}
Notice that $(\ups, \gamma_+\chi, \gamma_-\chib, E)$ and $(\upsb,
\gamma_-\chi, \gamma_+\chib, \Eb)$ transform among themselves.

We face a difficulty when we try to construct an $\CN = (2,2)$
supersymmetric action $S_{\Vt}$ for a twisted vector multiplet.  The
natural choice
\begin{equation}
  (\veps_\alpha\veps_\beta) S_{\Vt}
  = \CQb_\alpha \CQ_\beta
    \int \! \sqrt{h} \, \rmd^2x \Tr\Bigl(
    \kappab\gamma_{\hat3}\kappa + \frac{2i}{r} \rho\rhob\Bigr),
\end{equation}
which is manifestly invariant under $\CN = (2,2)$ supersymmetry, does
not work since its bosonic part
\begin{equation}
  \int \! \sqrt{h} \, \rmd^2x \Tr\Bigl(
  F_{\hat1\hat2}^2 + D^\mu\rhob D_\mu\rho - \frac{1}{r^2}\rhob\rho
  + \frac14[\rho,\rhob]^2 + C^2\Bigr)
\end{equation}
is not bounded from below, plagued by the negative potential term
$-\Tr\rhob\rho/r^2$.

\pagebreak

A sensible supersymmetric action $S_{\Ct}$ can be constructed for
twisted chiral multiplets.  It is convenient to shift
\begin{equation}
  E \to E - \frac{iq}{2r} \ups,
\end{equation}
which absorbs all the $q$-dependent terms from the supersymmetry
transformation \eqref{TCM-SUSY}.  Then the action can be written as
\begin{equation}
  (\veps_\alpha\veps_\beta) S_{\Ct}
  = \CQb_\alpha\CQ_\beta \int \! \sqrt{h} \, \rmd^2x
    \Bigl(\ups\Eb - \upsb E + \frac{i}{r} \upsb\ups\Bigr).
\end{equation}
Computing the supersymmetry variations we get
\begin{multline}
  S_{\Ct}
  = \int \! \sqrt{h} \, \rmd^2x \Bigl[
      \upsb\Bigl(-D_\mu^2+ \frac12\{\rho, \rhob\} + iC\Bigr)\ups
      + \Eb E \\
      - i\chi\Ds\chib
      + i\chi\Bigl(\gamma_+\kappab\ups+ \gamma_-\kappab\upsb
                   +\frac12\gamma_{\hat3}\rhob\chi\Bigr)
      + i\chib\Bigl(\gamma_-\kappa\ups + \gamma_+\kappa\upsb
                    + \frac12\gamma_{\hat3}\rho\chib\Bigr)\Bigr].
\end{multline}

For any gauge-invariant holomorphic function $\Wt(\ups)$ of twisted
chiral multiplet scalars $\ups^i$, the twisted F-term
\begin{equation}
  \label{SP}
  S_{\Wt} = i \int\! \sqrt{h} \, \rmd^2x
             \Bigl(E_{\Wt} + \Eb_{\Wt}
             + \frac{2}{r} \Im \Wt\Bigr)
\end{equation}
is invariant under $\CN = (2,2)$ supersymmetry.  Here $E_{\Wt}$ is the
auxiliary field of the twisted chiral multiplet $(\Wt, \chi_{\Wt},
\chib_{\Wt}, E_{\Wt})$, given by
\begin{equation}
  E_{\Wt} = E^i \del_i\Wt - (\chib^i\gamma_+\chi^j) \del_i\del_j\Wt.
\end{equation}

\providecommand{\href}[2]{#2}\begingroup\raggedright\endgroup
\end{document}